# A wide-band perfect light absorber at mid-wave infrared using multiplexed metal structures

Joshua Hendrickson,[1] Junpeng Guo,[2*] Boyang Zhang,[2] Walter Buchwald,[3] and Richard Soref[4]

[1]Sensors Directorate, Air Force Research Laboratory, Wright Patterson Air Force Base, Ohio 45433, USA
[2] Department of Electrical and Computer Engineering, University of Alabama in Huntsville, Huntsville, Alabama 35899, USA
[3]Solid State Scientific Corporation, Nashua, New Hampshire 03060, USA
[4]Department of Physics and Engineering Program, University of Massachusetts at Boston, Boston, Massachusetts 02125, USA
*Corresponding author: guoj@uah.edu

We experimentally demonstrate a wide band near perfect light absorber in the mid-wave infrared region using multiplexed plasmonic metal structures. The wide band near perfect light absorber is made of two different size gold metal squares multiplexed on a thin dielectric spacing layer on the top of a thick metal layer in each unit cell. We also fabricate regular non-multiplexed structure perfect light absorbers. The multiplexed structure IR absorber absorbs above 98% incident light over a much wider spectral band than the regular non-multiplexed structure perfect light absorbers in the mid-wave IR region.

Anomalous light absorption in metal structures was first observed a century ago by Wood [1]. The interest of light absorption in structured metals resurfaced in the 1960s, 1970s, and 1990s [2-8]. Today, it is well understood that anomalous light absorption in metal structures is due to the excitation of surface plasmon-polaritons (SPPs). Recently, perfect electromagnetic energy absorptions in structured metamaterials have been demonstrated in the gigahertz and terahertz regimes [9-10]. Perfect absorbers at optical frequencies have also been reported by several groups [11-16]. However, all the metamaterial perfect absorbers reported have very narrow spectral widths limited by the line-widths of the electromagnetic resonances in the structures. In many applications, it is desirable to have perfect absorption over broader spectral bands. Expansion of absorption band has been proposed using structures combining multiplexed subwavelength apertures [13], however, the proposed structure is polarization dependent and experimentally has not been demonstrated. In this paper, we report an experimental demonstration of a wide spectral band perfect absorber using a multiplexed surface plasmon resonance structure. In the multiplexed surface plasmon resonance structure, two gold metal squares are multiplexed in the unit cell of the periodic structure. The multiplexed plasmonic structure metamaterial, operating in the mid-wave infrared regime, near perfectly absorbs photons over a wider spectral range than previously reported.

Figure 1 (a) shows the regular non-multiplexed narrow band perfect light absorber structure. In this structure, gold thin film squares are patterned periodically on the top of a thin dielectric layer deposited on top of a thick gold metal layer. The thick metal layer is thick enough that no transmission can occur when light is incident from above the structure. Due to electromagnetic resonance in the metal-dielectric subwavelength structure, the effective impedance of the structured metamaterial surfaces can match the impedance of the vacuum; therefore reflections from the surface can be completely eliminated. Fig. 1 (b) shows the multiplexed perfect light absorber structure. The period of the multiplexed structure is the same as the period of the non-multiplexed perfect light absorber structure. However, in the multiplexed structure there are two metal squares of different sizes in the unit cell. Due to the size difference, two surface plasmon resonance modes at different frequencies can occur. In both of these structures, IR transparent magnesium fluoride ($MgF_2$) was chosen as the dielectric spacing layer. The periods of the unit cells are identical in both lateral dimensions to insure polarization independence at the normal incidence.

We designed the regular non-multiplexed perfect light absorbers and the multiplexed perfect light absorber in the mid-wave infrared region. The design simulations were carried out using a finite-difference time-domain software (Lumerical, Inc.). A Lorentz-Drude material model based on measurement data was used for the electric permittivity of the gold film [17]. The optical constant of the magnesium fluoride spacer was obtained from the reference [18]. The simulation domain has periodic boundary conditions in the lateral directions. In the directions of the propagation and reflection, the simulation domain is terminated with perfect matching layers. We first carried out geometrical parameter optimization for the gold square sizes and the $MgF_2$ spacing layer thickness to minimize the power reflections from the non-multiplexed and multiplexed structure absorbers for normal incidence. Fig. 2 shows the calculated power reflections from two non-multiplexed structure perfect absorbers and one multiplexed structure wide band perfect absorber. The dotted blue line is the power reflection from a regular non-multiplexed metal structure perfect absorber with the metal square size of 815 nm by 815 nm in the unit cell. The perfect absorption occurs at 3.347 μm. The reflectivity at this wavelength is 0.16%. The black dotted line in Fig. 2 is the power reflection from a regular non-multiplexed structure perfect light absorber with one 865 nm metal square in the unit cell. The perfect absorption occurs at the wavelength of 3.525 μm. The reflectivity at this wavelength is 0.06%. The solid red line curve in Fig. 2 is the power reflection from the multiplexed structure perfect absorber with two gold metal squares in the unit cell. The first gold square is 815 nm by 815 nm. The second gold square is 865 nm by 865 nm. The two gold metal squares are offset from each other by 210 nm in two lateral dimensions in the unit cell. In all three perfect absorber structures, the gold metal film thickness is 55 nm, and the magnesium fluoride layer thickness is 75 nm. The unit cells in the three perfect absorber structures repeat themselves with the same period of 2100 nm. It is seen clearly from Fig. 2 that the multiplexed structure perfect absorber achieves a much wider near perfect absorption band than the regular non-multiplexed structure absorbers.

We calculated the electric field intensity distribution in the middle plane inside the $MgF_2$ spacing layer (37.5 nm above the surface of the thick gold metal layer) in the unit cell of the multiplexed absorber structure. In the unit cell, the 815 nm gold square is on the bottom left and the 865 nm gold square is on top right. Fig. 3 (a) shows the electric field intensity distribution calculated at 3.35 µm, a wavelength on the short side of the wide absorption band. It can be seen that the 815 nm metal square is resonantly excited. Fig. 3 (b) shows the electric field intensity distribution at the wavelength of 3.395 µm, the center of the wide absorption band. At this wavelength, both metal squares exhibit strong local field enhancement. Fig. 3(c) shows the electric field intensity distribution the wavelength at 3.45 µm, on the long wavelength side of the wide absorption band. It can be seen that the 865 nm metal square exhibits strong field enhancement as expected. Finally, Fig. 3 (d) shows the electric field distribution at the wavelength of 4.0 µm, a wavelength outside of the absorption band. None of the plasmon resonance modes of two gold squares are excited. The electric field intensity distributions in Fig. 3 clearly show that the two plasmon resonance modes in the two gold squares cause the broadening of the absorption band.

We fabricated two non-multiplexed narrow band perfect light absorbers and one multiplexed structure wide band perfect light absorber in one fabrication process. The fabrication process started with a polished silicon wafer as the substrate. First, we deposited a 20 nm titanium adhesion layer on the silicon wafer substrate followed by a 250 nm thick gold layer. A 75 nm $MgF_2$ spacer layer was then deposited on the thick gold layer. After a wet chemical cleaning, PMMA 495k diluted with anisole at a 1:1 ratio was spin-coated onto the $MgF_2$ surface. Electron beam lithography was then used to define three different patterns: an array of 815 nm squares, an array of 865 nm squares, and an array of multiplexed squares of 815 nm and 865 nm in the unit cell. The periods of all three unit cell patterns are the same at 2100 nm. In the multiplexed structure, one gold metal square was offset from the other by 210 nm, horizontally and vertically. The overall size of each fabricated device is 100 µm². Each pattern is placed 5 mm from others in order to avoid possible cross talks during the subsequent measurement process. After development in a solution of MIBK/IPA, a 3 nm titanium adhesion layer and a 55 nm gold film were evaporated onto the patterned e-beam resist. A standard lift-off process was followed with acetone to remove the areas of unwanted material. Fig. 4 (a) shows a SEM image of the periodically arranged non-multiplexed 815 nm gold squares. Fig. 4 (b) shows a SEM image of the multiplexed structure where 815 nm and 865 nm gold metal squares are both included in the unit cell.

We measured the power reflection from the three fabricated perfect absorbers using a microscope coupled Fourier transform infrared spectrometer (Bruker Vertex 80V FTIR and Hyperion microscope). A 36 X reflecting objective lens was used in the microscope and the signal was detected with a mercury-cadmium-telluride (MCT) detector. The field of view of the microscope was reduced to 50 µm² with the aid of a set of apertures. This allows us to obtain measurements over a more uniform area of the 100 µm² patterns since proximity effects in the lithographic process may cause the patterned metal squares near the periphery of the devices to be undersized.

The power reflection measured from all three perfect absorbers versus the wavelength is plotted in Fig. 5. The dotted blue line is the power reflection from the regular non-multiplexed structure perfect absorber with one 815 nm gold square in the unit cell. The device has near perfect absorption of 96% at 3.36 µm wavelength. The dotted black line is the power reflection from the regular non-multiplexed structure perfect absorber with one 865 nm gold square in the unit cell. This device has near perfect absorption of 96.7 % at 3.55 µm wavelength. It has a longer near perfect absorption wavelength because of the larger size of the metal square in the unit cell. The solid red line in Fig. 5 is the measured power reflectivity from the multiplexed structure perfect absorber with 815 nm and 865 nm gold metal squares in the unit cell. The multiplexed structure's absorption reaches above 98% over a wide spectral band centered at 3.45 µm wavelength. It can be seen that the multiplexed structure's absorption band has been expanded significantly due to the two gold metal squares of different sizes in the unit cell.

In summary, we have experimentally demonstrated a wide band perfect light absorber using multiplexed metal structures with two plasmon resonance modes in the unit cell. The multiplexed structure metamaterial surface near perfectly absorbs above 98% IR light over a wide spectral band centered at the mid-IR wavelength of 3.45 µm. Our experimental results agree well with numerical simulation results. The demonstrated multiplexed structure wide band perfect absorption concept can be applied for future applications in enhanced optical signal sensing and imaging.

Prof. Junpeng Guo acknowledges the support from the ASEE-Air Force Summer Faculty Research Fellowship Program. Joshua Hendrickson acknowledges the support from the Air Force Office of Scientific Research (AFOSR) through the program LRIR 10RY04COR. This work was also supported by the National Science Foundation (NSF) through grant NSF No. 0814103 and the National Aeronautics and Space Administration (NASA) through contract NNX07 AL52A. .

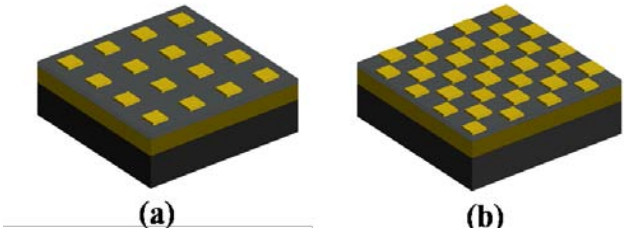

Fig. 1. (Color online) (a) The regular non-multiplexed perfect light absorber structure; (b) the multiplexed perfect light absorber structure.

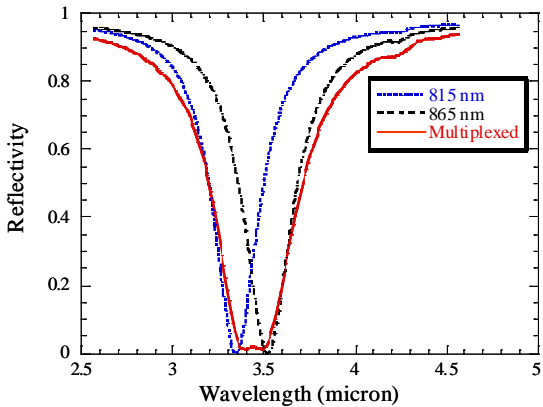

Fig. 2. (Color online) Calculated power reflections from three perfect light absorbers. The dotted blue line is the power reflection from the perfect absorber structure with one 815 nm gold square in the unit cell. The black dotted line is the power reflection from the perfect absorber with one 865 nm metal square in the unit cell. The red line curve is the power reflection from the multiplexed structure perfect absorber with two gold metal squares in the unit cell.

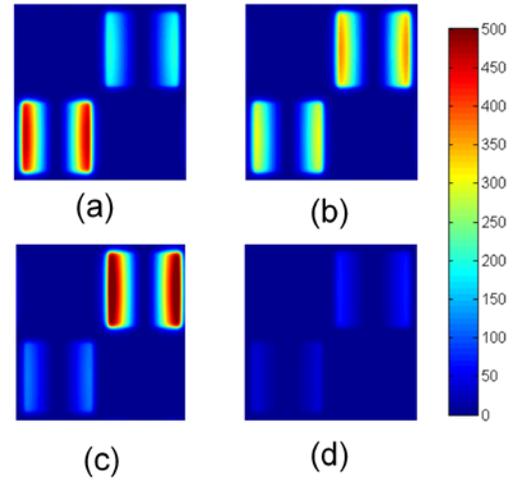

Fig. 3. (Color online) Electric field intensity distributions in the center plane of the $MgF_2$ spacing layer in the unit cell of the multiplexed absorber structure: (a) electric field intensity distribution at 3.35 μm wavelength; (b) electric field intensity distribution at 3.395 μm wavelength in the center of the wide absorption band; (c) electric field intensity distribution at 3.45 μm wavelength; (d) electric field intensity distribution at 4.0 μm which is outside of the absorption band.

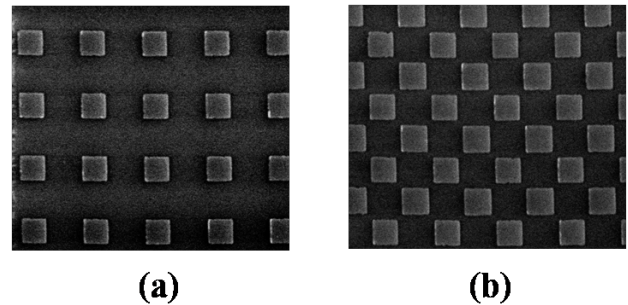

Fig. 4. (Color online) Scanning electron micrograph (SEM) images of the gold squares in the fabricated perfect light absorbers: (a) the regular non-multiplexed structure including one 815 nm gold square in the unit cell; b) the wide band multiplexed structure including 815 nm and 865 nm gold squares in the unit cell.

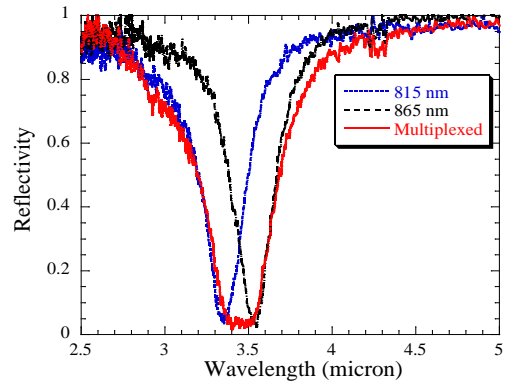

Fig. 5. (Color online) The measured power reflection versus the wavelength from the multiplexed gold metal structure perfect absorber (red line), the non-multiplexed structure absorber of 815 x 815 nm² metal square (dotted blue line), the non-multiplexed structure absorber of 865 x 865 nm² metal square (dotted black line) in the unit cells.